\documentclass{jpsj3}
\usepackage{txfonts}

\title{Local Representation of $N$-body Coulomb Energy with Path Integrals}

\author{\name{Takanori Sugihara}$^1$\thanks{E-mail: t-sugihara@aist.go.jp}, \name{Junichi Higo}$^2$, and \name{Haruki Nakamura}$^2$}
\inst{$^1$Japan Biological Informatics Consortium (JBIC), 2-4-32 Aomi, Koto-ku, Tokyo, 135-0064, Japan \\
$^2$Institute for Protein Research, Osaka University, 3-2 Yamadaoka, Suita, Osaka 565-0871, Japan} 

\abst{We represent $N$-body Coulomb energy in a localized form 
to achieve massive parallelism. 
It is a well-known fact that Green's functions can be written 
as path integrals of field theory. 
Since two-body Coulomb potential is a Green's function of Poisson equations, 
it reduces to a path integral of free scalar field theory 
with three spatial dimensions. 
This means that $N$-body one also reduces to a path integral. 
We discretize real space with a cubic lattice 
and evaluate the obtained multiple integrals approximately 
with the Markov-chain Monte Carlo method. }

\begin{document}
\maketitle

Molecular dynamics (MD) is a numerical method to describe 
motion of $N$ interacting particles based on 
Newton's classical equations of motion \cite{alder}. 
It has been used in various fields such as structure prediction 
of protein \cite{protein}, in silico screening \cite{insilico}, 
and material design. 
In MD of atoms, computationally the heaviest part is calculations of 
$N$-body Coulomb interaction, which explodes with operations 
proportional to O($N^2$). 
To propel research and development based on MD, 
we need to decrease turnaround time of computation 
by processing large amount of arithmetic operations 
in terms of state-of-the-art high-performance supercomputers. 

There have been many efforts to decrease 
the number of operations to calculate 
$N$-body Coulomb interaction, such as 
Ewald \cite{ewald}, fast multipole \cite{fmm}, 
and Wolf \cite{wolf} methods. 
The methods beautifully treat the non-locality 
associated with the long-range Coulomb interactions 
by dividing those into two: short and long-range regions. 
However, to parallelize relatively large molecules 
based on the methods, 
one has to face again the difficulty due to non-locality. 
There is no choice but to accept 
large data communications and load imbalance 
for massive parallelism. 
We cannot conduct parallel computations 
without exchanging the whole $N$ or 
a partial sets of coordinates among computation nodes. 

One can detour around the non-locality associated with 
the Coulomb's long-range tails 
by parallelizing Markov-chain generation processes \cite{sugi}. 
However, it is a sampling method to obtain statistical 
quantities such as free energy and does not simulate 
actual time-dependent conformations of molecules. 
In addition, it does not work for larger systems 
due to limitations on available memory size. 
A computation node is too small to treat 
a huge molecular system 
made of millions or more atoms. 
For a fundamental solution to the hard problem, 
we still need to find a new idea 
to achieve locality and parallelism. 

In this letter, we rewrite a summation 
of $N$-body Coulomb energy in a local form 
based on path integral formulation 
of field theory \cite{path}. 
Since everything is modeled in latticized 
real space, spatial decomposition is possible 
and gives rise to massive parallelism.

The method can take open boundary conditions. 
Therefore, it can treat isolated systems. 
Also other boundary conditions such as 
periodic and fixed ones are available if necessary. 
Any other types of boundary conditions work 
if those can be implemented in a field theoretic manner. 
The method has no conditions 
associated with the total charge of atoms. 
Therefore, it can be applied to classical gravity systems. 

In this method, the number of operations necessary 
to calculate $N$-body Coulomb energy is O($N$). 
The dominant operations are 
fused multiply-add ones, 
which can be processed in parallel. 
As shown later, the method is suited to 
relatively large-$N$ systems 
because it has good weak scaling property 
when a parallel computer is applied. 
As for data communication among computation nodes, 
one needs to perform global communication once 
every $N$-body Coulomb energy summation. 
This is because one needs to collect results 
from all the computation nodes and sum up 
those after all necessary calculation are done.

Let us move onto theoretical details of the method. 
Coulomb energy among $N$ charged particles is 
a summation of two-body ones:
\begin{equation}
E_N = 
\frac{1}{4\pi}\sum_{i \neq j}^N
 \frac{q_i q_j}{|r_i-r_j|}, 
\label{s1}
\end{equation}
where $r_i$ and $q_i$ represent a vector 
in three-dimensional real space $R^3$ and 
an electric charge of the $i$-th particle, respectively. 
In general, Greens' functions can be written 
as path integrals based on the method of 
generating functional in field theory \cite{abers}. 
Two-body Coulomb potential is inverse of 
Laplacian $\Delta$ 
\begin{equation}
\Delta \frac{1}{|r|} = -4\pi \delta^3(r).
\end{equation}
Therefore, we can represent it in a path-integral form:
\begin{equation}
\frac{1}{4\pi |r_1-r_2|}
=
Z[0]^{-1}
\left.
\frac{\delta}{\delta J(r_1)}
\frac{\delta}{\delta J(r_2)} Z[J]\right|_{J=0}, 
\label{2body}
\end{equation}
where
\begin{equation}
Z[J] = \int {\cal D}\phi e^{-H[J]}, 
\label{part}
\end{equation}
and
\begin{equation}
H[J] = \int dr^3 \left(
     -\frac{1}{2} \phi(r) \Delta \phi(r)
     -J(r) \phi(r) \right).
\end{equation}
${\cal D}\phi$ is Feynman's path integral measure 
\begin{equation}
{\cal D}\phi \propto
\prod_{r\in R^3} d\phi(r). 
\end{equation}
In Eq. (\ref{part}), we can integrate out 
the functional variable $\phi(r)$ 
because the integrand is Gaussian 
\begin{equation}
Z[J] = Z[0] \exp \left[ \frac{1}{2} 
\int dr_1^3 dr_2^3 J(r_1) \frac{1}{4\pi |r_1-r_2|} J(r_2)
\right], 
\end{equation}
which gives Eq. (\ref{2body}). 
Then, we have 
\begin{equation}
\frac{1}{4\pi |r_1-r_2|}
=
\frac{\int {\cal D} \phi \, 
\phi(r_1)\phi(r_2)
e^{-H[0]}}
{\int {\cal D} \phi \, e^{-H[0]}}. 
\label{path}
\end{equation}
This is a building block of our new method. 

We would like to evaluate the right hand side of Eq. (\ref{path}) 
numerically. For this purpose, we discretize 
three-dimensional space $R^3$ with a cubic lattice $a^3$, 
where $a$ is a spacing between two lattice sites \cite{creutz}. 
Hereafter, we omit lattice spacing $a$ for brevity. 
In the latticized real space $L^3$, where $R=aL$, 
the counter part of Eq. (\ref{path}) is given by 
\begin{equation}
\frac{1}{4\pi|m-n|}
\approx
\frac{\displaystyle\int {\cal D}_L \phi \, 
\phi_m \phi_n
e^{-H_{L}}}
{\displaystyle\int {\cal D}_L \phi \, e^{-H_L}}, 
\label{proplat}
\end{equation}
where
\begin{equation}
H_L = \frac{1}{2} \sum_{\langle m,n \rangle} (\phi_m - \phi_n)^2, 
\quad
{\cal D}_L \phi \propto
\prod_{n \in L^3} d\phi_n. 
\end{equation}
We have replaced the derivative with the first-order difference. 
The lowercase letter $m$ represents a position 
$(m_x,m_y,m_z)$ in the lattice $L^3$, 
where $r_x=m_x a$ and so on. 
The brackets $\langle m,n \rangle$ represents the set of 
all nearest-neighbor pairs of lattice sites.

\begin{figure}
\begin{center}
\includegraphics[width=90mm]{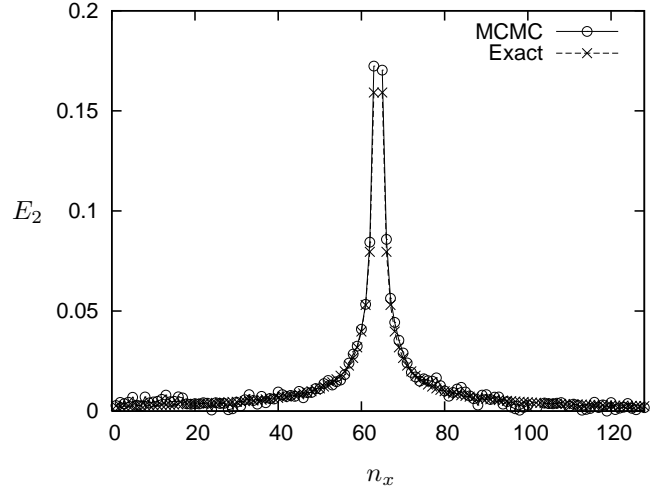}
\end{center}
\caption{The two-body Coulomb energy $E_2$ is compared 
between MCMC (circles with solid line) and 
the exact (crosses with dashed line) results 
on a lattice $L=128$.
$E_2$ is plotted as a function of $n_x$ 
with $m_x=L/2$, $m_y = n_y = L/2$, 
and $m_z = n_z = L/2$ fixed. 
MCMC calculations have been done 
with the sample number $M=10^4$. 
\label{c}}
\end{figure}

For two-body Coulomb energy in Eq. (\ref{proplat}), 
it is much easier to evaluate the left hand side than right. 
However, for $N$-body Coulomb energy with relatively large $N$, 
the path-integral form in the right hand side may be useful: 
\begin{eqnarray}
E_N
&\approx&
\frac{\displaystyle\int {\cal D}_L \phi \, 
\displaystyle
\Bigg[
\Bigg(\sum_{i=1}^N q_i \phi_{n_i} \Bigg)^2
-\sum_{i=1}^N \left(q_i \phi_{n_i}\right)^2 
\Bigg] \,
e^{-H_L}}
{\displaystyle\int {\cal D}_L \phi \, e^{-H_L}}. \nonumber\\
\label{nsum}
\end{eqnarray}
We are going to evaluate the multiple integral 
in the right hand side with Markov chain Monte Carlo (MCMC) method. 
Concretely, we use the heathbath algorithm \cite{glauber}, 
which may be more efficient than Metropolis \cite{metropolis} 
because acceptance-rejection process is not necessary and 
therefore the acceptance ratio is unity. 
In the heatbath algorithm, a new configuration $C'$ is 
generated independent of an old one $C$ so that 
probability distribution $P$ satisfies 
the following relation. 
\begin{equation}
P(C\to C') \propto e^{-H_L(C')}, 
\end{equation}
which is a sufficient condition of the detailed balance. 
According to this, a lattice variable $\phi_n$ is updated with 
\begin{equation}
r=
\frac{\displaystyle\int_{-\infty}^{\phi_n} 
d\varphi \, e^{-\frac{1}{2} 
\sum_m \left( \varphi - \phi_m \right)^2}}
{\displaystyle\int_{-\infty}^\infty d\varphi \, 
e^{-\frac{1}{2} \sum_m \left( \varphi - \phi_m \right)^2}}, 
\label{hb1}
\end{equation}
where $r$ is a uniform random number $0\le r \le 1$. 
By inverting Eq. (\ref{hb1}), we have 
\begin{equation}
\phi_n =
-\frac{1}{\sqrt{3}} {\rm erfc}^{-1}(2r)
+\frac{1}{6} \sum_m \phi_m, 
\label{phi}
\end{equation}
where the summation $m$ is taken for 
all the nearest-neighbor sites of $n$. 
One can generate a new configuration $C'$ by substituting 
a uniform random number $r$ to Eq. (\ref{phi}). 
(One has to modify Eq. (\ref{phi}) slightly at the boundary 
when open boundary conditions are imposed 
because there is no lattice sites outside the lattice.) 
All the lattice variables are updated sequentially 
in terms of Eq. (\ref{phi}). 
Once a Monte-Carlo sample set $\{ \phi_n^{(j)}: j=1,2,\dots,M \}$ is obtained, 
Eq. (\ref{nsum}) can be approximated as 
\begin{equation}
E_N \approx
\frac{1}{M} \sum_{j=1}^M 
\Bigg[
\Bigg( \sum_{i=1}^N q_i \phi_{n_i}^{(j)} \Bigg)^2
-\sum_{i=1}^N \left(q_i \phi_{n_i}^{(j)}\right)^2
\Bigg],
\label{nsum2}
\end{equation}
where $M$ indicates the number of Monte-Carlo samples. 
The number of arithmetic operations necessary to 
evaluate the right hand side of Eq. (\ref{nsum2}) is O($N$) 
for a fixed $M$.
As $N$ goes large with $M$ fixed, there is a some point 
where Eq. (\ref{nsum}) is more advantageous than (\ref{s1}) 
because direct evaluation of the latter requires 
operations of O($N^2$). 
One can adjust $M$ to control statistical errors 
according to available computing resources. 
In addition, the right hand side of Eq. (\ref{nsum2}) has 
massive parallelism 
because each term contained in the summations 
can be evaluated independently 
without any information related to other particles.

In order for the method to be practical, 
it needs to reproduce two-body Coulomb 
energy accurately as much as possible because 
it is a building block of $N$-body one. 
Let us compare two-body Coulomb energy 
between MCMC and exact results, 
which are calculated 
by making use of Eqs. (\ref{nsum2}) and (\ref{s1}), respectively. 
We set $N=2$ and $q_i =1$ and choose open boundary conditions. 
We put the first particle at the center of a latticized real space, 
where a lattice site $m$ takes values of 
$m_x=L/2$, $m_y =L/2$, and $m_z=L/2$. 
Then we move the second particle along a straight line 
$n_y=L/2$ and $n_z=L/2$ and measure energy between two. 
Figure \ref{c} plots two-body Coulomb energy 
as a function of $n_x$ for $L=128$. 
MCMC and exact results are plotted with open circles 
and crosses, respectively. 
As for MCMC, the number of samples is $M=10^4$. 
Samples are collected controlling correlations 
among configurations as usual. 
Actually, we take one sample every $1000$ sweeps of 
$L^3$ lattice sites with Eq. (\ref{phi}). 
We can say that the obtained MCMC statistical average 
with $M=10^4$ is good approximation to the exact one 
when the lattice size is $L=128$.

\begin{figure}
\begin{center}
\includegraphics[width=90mm]{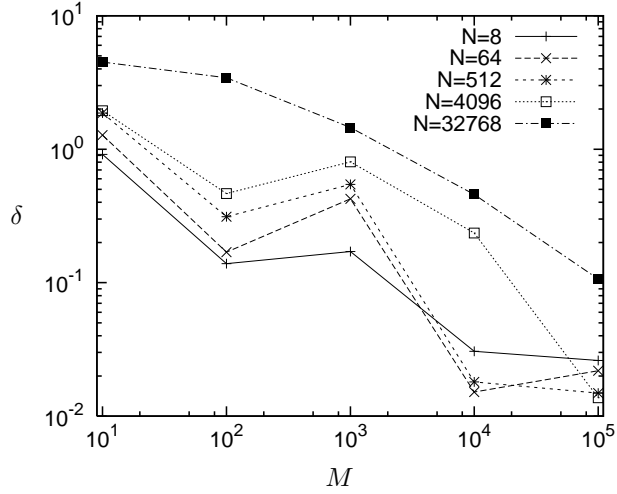}
\end{center}
\caption{The difference $\delta$ 
between MCMC and the exact results divided by the exact 
is plotted as a function of the number of MCMC samples $M$ 
for various atom numbers $N=8,64,512,4096$ and $32768$ 
on a lattice $L=128$ with open boundary conditions. 
\label{conv128}}
\end{figure}

\begin{figure}
\begin{center}
\includegraphics[width=90mm]{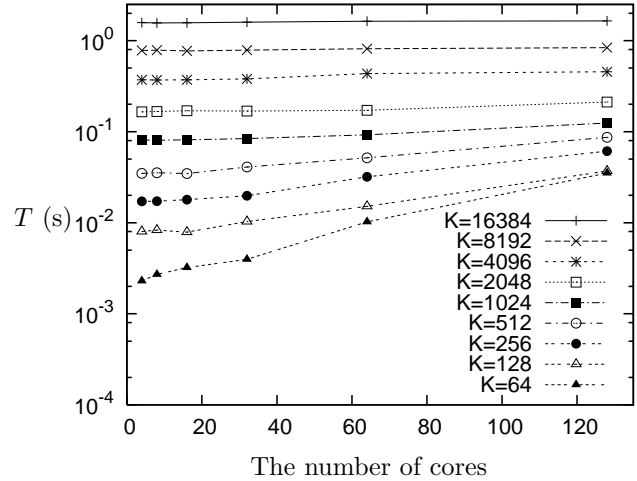}
\end{center}
\caption{Weak-scaling property of the method is shown. 
Elapsed time $T$ is plotted using generated $M=10^4$ samples 
as a function of the number of cores for various unit problem size 
$K=64,128,256,512,1024,2048,4096,8192$ and $16384$. 
For larger $K$, the plotted lines are almost flat. 
For smaller $K$, slightly increase due to use of 
the collective communication {\tt mpi\_reduce}. 
\label{weak}}
\end{figure}

In order to check accuracy of the method 
in many-body calculations, 
we measure errors associated with $N$-body Coulomb energy. 
Let us consider the following quantity: 
\begin{equation}
\delta \equiv 
\frac{\left|({\rm MCMC}) - ({\rm Exact})\right|}
{\left|({\rm Exact})\right|}, 
\label{aver}
\end{equation}
which is the difference between MCMC and exact results of 
$N$-body Coulomb energy divided by the exact. 
In Fig. \ref{conv128}, Eq. (\ref{aver}) is plotted 
as a function of the number of Monte Carlo samples $M$ 
for various numbers of particles $N=2^3$, $4^3$, $8^3$, $16^3$, 
and $32^3$ on the lattice $L=128$ with open boundary conditions. 
We choose all charges are $q_i = 1$ as before 
to show that the method works also in general cases 
such as gravity systems. 
For simplicity, we put $N$ charged particles separately 
on lattice sites in a region 
$L/2-N^{1/3}/2+1 \leq n_i \leq L/2+N^{1/3}/2$ for $i=x,y,z$, 
which forms a cube. 
Errors associated with $N$-body Coulomb energy 
decreases as $M$ increases, 
which is consistent with a well-known fact 
that errors associated with Monte-Carlo sampling 
are proportional to $1/\sqrt{M}$. 
We admit that the open boundary conditions affect 
accuracy associated with larger systems 
such as $N=4096$ and $32768$.  
This means that, 
in many-body calculations with large $N$, 
we need to take a sufficiently large $L$ 
to produce good approximate results 
avoiding boundary effects.

The most remarkable feature of the proposed method is 
scalability for large $N$ in parallel computation. 
In order to demonstrate it, we parallelize calculations 
of Eq. (\ref{nsum2}) 
and check their weak-scaling property. 
We consider a rectangular parallelepiped 
with $K$ lattice sites 
that is processed by one computation core. 
We put a particle on every lattice site and 
fix the number of lattice sites (i.e. particles). 
We call $K$ the unit problem size. 
Therefore, the total number of particles is 
a product of the unit problems size $K$ and 
the number of computation cores. 
When all the calculations are done, the results are 
collected from all the computation nodes and summed up 
to calculate Eq. (\ref{nsum2}) 
in terms of the MPI collective communication function 
{\tt mpi\_reduce} \cite{mpi}. 
Figure \ref{weak} shows weak-scaling property of the method, 
where elapsed time $T$ is plotted using generated $M=10^4$ samples 
as a function of the number of cores 
for various unit problem size $K$. 
That is, the number of cores is changed 
with the unit problem size $K$ fixed. 
Ideally, plotted lines should be flat because 
the number of operations processed by each core is 
identical and there is no data communication among cores. 
For larger $K$, we can say that 
the actual weak-scaling performance is very close to the ideal. 
The method can treat very large systems 
composed of millions or more particles 
with almost ideal parallel performance. 
On the other hand, for smaller $K$, 
we see that the plotted lines are 
slightly increasing because of the use of 
the collective communication 
in actual implementation of the method. 
This means that 
weak-scaling property of the method depends on 
the performance of the collective communication 
especially when $K$ is small.

We have carried the numerical calculations 
on a PC cluster with Intel Xeon X5460 (3.16 GHz) processors 
and Infiniband inter-node connections. 
We have generated executable binary codes 
with Intel Fortran compilers.

It costs a lot to generate 
MCMC sample set $\{ \phi_n^{(j)} \}$ because 
the number of necessary operations is proportional to $ML^3$. 
To use larger $L$ for accuracy, one needs longer CPU time and 
larger storage to generate and store MCMC samples, 
respectively. 
However, a sample set can be reused 
once generated appropriately 
because it depends only on lattice size $L$ 
but not positions of charged particles $n_i$.

To decrease errors associated with the rapidly 
increasing core part of the two-body Coulomb potential, 
one could calculate close particles directly. 
It does not require large data communications 
to do that if most of close particles are 
contained in the same or neighbor computation nodes.

Although we have put particles on lattice sites 
just for simplicity, that is not necessary. 
To put particles arbitrarily, 
one can model coordinate values of every particle 
in terms of the nearest eight lattice sites 
in three-dimensional lattice space, for example. 
Also one can evaluate forces among particles 
with interpolation in a similar way. 

In this letter, we have chosen $q_i=1$ for 
all the particles. Of course one can produce 
many-body energy with arbitrary charges 
in the same accuracy as the two-body energy. 
Errors of many-body energy comes from two-body one 
because $N$-body one is a superposition of two-body. 
It is important to take larger 
$L$ for higher accuracy.

The proposed parallelization algorithm is quite simple 
because it is just a combination of fused multiple-add 
operations and reduction communications. 
It does not require any other expensive operations 
such as division. 
We can maximize the performance of the method 
if we have computers customized for processing 
those operations and communications.

\begin{acknowledgment}

This work was supported by grants from the METI (Ministry of Economy, Trade and Industry)
Project to build infrastructure for creating next-generation drugs for personalized medicine.

\end{acknowledgment}


\begin{thebibliography}{9}

\bibitem{alder}
B.~J.~Alder and T.~E.~Wainwright, 
J. Chem. Phys. {\bf 27}, 1208 (1957). 

\bibitem{protein}
{\it Protein Folding}, edited by T.~E.~Creighton 
(Freeman, New York, 1992). 

\bibitem{insilico}
I.~D.~Kuntz, J.~M.~Blaney, S.~J.~Oatley, and R.~Langridge, T.~E.~Ferrin, 
J. Mol. Biol. {\bf 161}, 269 (1982). 

\bibitem{ewald}
P.~P.~Ewald, Ann. Phys. {\bf 64}, 253 (1921); 
T.~Darden, D.~York, and L.~Pedersen, 
J. Chem. Phys. {\bf 98}, 10089 (1993). 

\bibitem{fmm}
H.~Q.~Ding, N.~Karasawa, and W.~A.~Goddard III, 
J. Chem. Phys., {\bf 97}, 4309 (1992); 
L.~Greengard and V.~Rokhlin, 
J. Comput. Phys. {\bf 73}, 325 (1987). 

\bibitem{wolf}
D.~Wolf, P.~Keblinski, S.~R.~Phillpot, and J.~Eggebrecht, 
J. Chem. Phys., {\bf 110}, 8254 (1999); 
C.~J.~Fennell and J.~D.~Gezelter, 
J. Chem. Phys., {\bf 124}, 234104 (2006); 
I.~Fukuda, Y.~Yonezawa, and H.~Nakamura, 
J. Phys. Soc. Jpn., {\bf 77}, 114301 (2008). 

\bibitem{sugi}
T.~Sugihara, J.~Higo, and H.~Nakamura, J. Phys. Soc. Jpn. {\bf 78}, 074003 (2009); 
J.~Higo, N.~Kamiya, T.~Sugihara, Y.~Yonezawa, and H.~Nakamura, Chem. Phys. Lett. {\bf 473}, 326 (2009);
J.~Ikebe, K.~Umezawa, N.~Kamiya, T.~Sugihara, Y.~Yonezawa, Y.~Takano, H.~Nakamura, and J.~Higo, J. Comp. Chem. {\bf 32}, 1286 (2011).

\bibitem{path}
R.~P.~Feynman and A.~R.~Hibbs, 
{\it Quantum Mechanics and Path Integrals} (McGraw-Hill, 1965). 

\bibitem{abers}
E.~Abers and B.~W.~Lee, Phys. Reports {\bf 9C}, 1 (1973). 

\bibitem{creutz}
M.~Creutz, {\it Quarks, Gluons, and Lattices} (Cambridge, 1983). 

\bibitem{glauber}
R.~Glauber, J. Math. Phys. {\bf 4}, 294 (1963). 

\bibitem{metropolis}
N.~Metropolis, A.~W.~Rosenbluth, M.~N.~Rosenbluth, and A.~H.~Teller: 
J. Chem. Phys. {\bf 21} (1953) 1087.

\bibitem{mpi}
http://www.mpi-forum.org/

\end{thebibliography}
\end{document}